\def\den{\hbox{den}}

\documentstyle[12pt]{article}
\begin{document}
\begin{titlepage} \vspace{0.2in} \begin{flushright}
MITH-95/20 \\ \end{flushright} \vspace*{0.8cm}
\begin{center} {\LARGE \bf  
Getting Around the Nielsen-Ninomiya Theorem,\\
towards the Rome Approach\\} \vspace*{0.5cm}
{\bf Giuliano Preparata and She-Sheng Xue$^{(a)}$}\\ \vspace*{0.7cm}
Dipartimento di Fisica, Universit\`a di Milano.\\
INFN - Section of Milan, Via Celoria 16, Milan, Italy\\ \vspace*{0.9cm}
{\bf   Abstract   \\ } \end{center} \indent

The ``no-go'' theorem of Nielsen and Ninomiya has been the most tenacious
obstacle against the construction of a chiral gauge theory with reasonable low
energy spectrum, couplings and anomaly. In this paper we construct a model
which supplements the usual (bilinear in the Fermi fields) lagrangian with
quadrilinear fermionic terms. We show that in a certain region of the parameter
space the difficulties of the ``no-go'' theorem may be overcome, and a
``renormalized'' perturbative strategy can be carried out, akin to the one
followed in the Rome Approach (RA), whose counterterms are forced to be gauge
invariant.

\vfill \begin{flushleft} Milano,16 October 1995   \vspace*{1.2cm} \\
\noindent{\rule[-.3cm]{5cm}{.02cm}} \\
\vspace*{0.2cm} \hspace*{0.5cm} ${}^{a)}$
E-mail address: xue@milano.infn.it\end{flushleft} \end{titlepage}

\noindent
{\bf 1.\hskip0.3cm}
The problem of defining a quantum chiral gauge field theory on a lattice lies 
at the roots of the Standard Model (SM). For even if space-time, the physical 
arena of the SM, is a continuum (as modeled in the generally accepted  SM, 
which we shall call Continuum Standard Model (CSM)) well beyond the 
Planck length ($a_p \simeq 10^{-33}$ cm), the very process of 
``renormalizing'' the theory requires its definition on a space-time, like a 
discrete lattice, where the energy-momentum spectrum of the quantum 
fluctuations is bounded from above. When this can be done, i.e. when we can 
define our chiral gauge theory on a lattice with arbitrarily small lattice 
constant $a$, the continuum limit is simply defined by the limit $a 
\rightarrow 0$.

It seems to us that this is the (essentially unique) way to give any 
sense to a Quantum Field Theory (QFT), and this for two reasons: one physical, 
the other mathematical. From the standpoint of physics the assumption 
of a mathematical continuum as the space-time basis of a generic QFT 
totally neglects the possible effects of the physics of Quantum 
Gravity in determining (according to the seminal ideas of Bernhard 
Riemann) the ``fine-structure'' of the (geometry of the) set of 
physical events. As for mathematics itself a continuum QFT contains 
{\em actual infinities}, a nightmare sternly but pointlessy denounced by 
Dirac, which can be possibly exorcized through a limiting procedure upon 
theories defined on lattices  whose constants decrease to zero. 

It is for the reasons just discussed that a most simple and general 
theorem \cite {nn}, found by H.B.Nielsen and M.Ninomiya at the beginning 
of the Eighties, appears as a most disappointing obstacle \cite{p} to giving 
a definite quantum field theoretical sense to the Standard Model itself, whose 
electroweak sector comprises precisely a quantum chiral gauge field 
theory. Prompted by the clear physical indication coming from the 
stunning phenomenological successes of the SM, L.Maiani and 
collaborators in Rome have elaborated a most reasonable philosophy to 
get around the unfavourable consequences of Nielsen and Ninomiya's ``no-go 
theorem'', without actually explicitly constructing a lattice QFT that 
evades the ``no-go theorem'', and at the same time yields in the 
continuum limit ($a\rightarrow 0$) the correct phenomenology. The 
analysis at one-loop level carried out by this group \cite{rome} confirms 
the soundness of their point of view. However it is clear that a positive 
answer to the question whether the favourable signs now available 
will turn the Rome Approach (as it has been called, RA) into a 
consistent fool-proof strategy to calculate the chiral gauge theory in 
the continuum limit, can be given if we are able to explicitly 
construct a consistent lattice chiral gauge theory whose continuum 
limit ($a\rightarrow 0$) is just the ``target theory'' of the RA.

It is worthwhile recalling that in the last few years we have tried to develop 
a research program 
centered on the formulation of the SM on a Planck lattice, a lattice 
whose constant {\em a} is just the Planck length $a_p$. The main physics 
motivation of our program is the expectation \cite{wheeler} that the 
violent quantum fluctuations of Quantum Gravity (QG) at the Planck 
scale do give rise to a discrete space-time structure, that can be 
usefully modeled by a 4-dimensional lattice, we have christened such 
attempt the Planck Lattice Standard Model (PLSM). A crucial aspect of 
the PLSM is the basic chiral nature of the electroweak sector, thus a 
basic prerequisite for its success is our ability to find an explicit 
formulation of a chiral gauge theory that
\begin {enumerate}
\item removes the ``doubling'' of the Weyl fermions and retains the 
correct anomaly;
\item maintains in the low-energy spectrum the basic chiral coupling to 
the gauge field;
\item realizes a ``renormalized'' perturbative strategy akin to the one known
as the Rome Approach without gauge-variant counterterms and pushing the gauge 
bosons' masses to the Planck mass, $m_p$.
\end {enumerate}

In this paper we shall introduce an explicit chiral lattice QFT model and show 
that in a certain region of the relevant parameter space, 
the ``physical wedge'', it satisfies the above three requirements. However, 
we should like to point out that we do not make any 
attempt\footnote {This will be reported in a future publication.} to 
formulate a fully relativistic PLSM of quark and lepton families in 
terms of our chiral model, thus we can keep its structure as simple as 
possible.

The model we shall analyse below is the simplest $SU(2)$ chiral gauge 
theory, which in order to avoid the sanctions of the ``no-go 
theorem'' will be supplemented with quadrilinear Fermi interactions 
terms\cite{ep}\cite{xue}, which we shall call the Nambu-Jona Lasinio (NJL) terms. 
Thus the fermionic part of the action on the lattice (of lattice 
constant $a$) is written as (repeated indices are summed over)
\begin{eqnarray}
S_F&&\!=\!{1\over 2a}\sum_n a^4\bar\psi^i_L(n)\gamma_\mu D^\mu_{ij}\psi^j_L+ 
{1\over 
2a}\sum_n a^4\bar\psi^l_R(n)\gamma_\mu\partial^\mu\psi^l_R(n)\nonumber\\&&+\!
g_1 
a^2\!\sum_n\!a^4\bar\psi^i_L(n)\cdot\psi^l_R(n)\bar\psi^l_R(n)\cdot\psi^i_L(n)\nonumber
\\&&\!+\!g_2a^2\!\sum_{n\mu\nu}\!a^4\bar\psi^i_L(n)\cdot 
(\partial^2_\mu\psi^l_R(n))
(\partial^2_\nu\bar\psi^l_R(n))\cdot\psi^i_L(n),
\label {action}
\end{eqnarray}
where $\psi^i_L(n)$ $(i=1,2)$ is the $SU(2)_L$ gauged doublet, 
$\psi^l_R(n)$ $(l=1,2)$ is a doublet, invariant under the transformations 
of $SU(2)_L$; both $\psi^i_L$ and $\psi^l_R$ are two-component Weyl 
fermions. In the second NJL-term
\begin{equation} 
\partial^2_\mu\psi^l_R(n)=\psi^l_R(n+\mu)+\psi^l_R(n-\mu)-2\psi^l_R(n), 
\label{d2}
\end{equation}
in the continuum limit this term is a dimension-10 operator which is 
relevant only for ``doublers'', i.e. for those field modes whose momenta 
\begin{equation}
p=\tilde p+\Pi_A,
\label{doublers}
\end{equation}
$\Pi_A$ being one of the fifteen Brillouin momenta and 
$|\tilde p|\ll {\pi\over a}$. Note that in addition to the 
$SU(2)_L$ chiral gauge invariance, and the global $SU(2)_L\times 
SU(2)_R$ symmetry, the fermion action (\ref{action}) possesses when $g_1=0$ an 
exact $\psi_R$-shift symmetry \cite{gp}:
\begin{equation}
\psi^l_R(n)\rightarrow \psi^l_R(n)+const.
\label{shift}
\end{equation}

\vskip1cm
\noindent
{\bf 2.\hskip0.3cm}
Our goal now is to seek a possible region of the $(g_1,g_2,g)$-space 
($g$ is the $SU(2)$ gauge coupling constant) where there ``lives'' a 
theory that obeys {\em the requirements $1.-3.$ above.} To start with we 
shall assume that the gauge coupling is a good perturbative coupling, 
thus for the time being we shall take $g \rightarrow 0$.

In the ($g_1,g_2$)-plane, in the weak coupling limit $g_1,g_2 \ll 1$, 
as indicated in fig.1, the action (\ref{action}) defines a $SU(2)_L \times 
SU(2)_R$ chiral continuum theory with a ``doubled'' fermion 
spectrum, {\em a theory that violates requirement $1. $}. On increasing 
$g_1$, while keeping $g_2$ in the perturbative region $g_2 \ll 1$, 
following the analysis and the discussion of Eichten and Preskill (EP) 
\cite{ep}, we shall hit a critical value $g^c_1$, above which the 
theory undergoes a phase transition. Using a strong coupling expansion 
in $g_1$ $ (g_1 > g^c_1)$ we can show 
that the fields $\psi^i_L$ and $\psi^l_R$ appearing in 
(\ref{action}) 
pair up with the composite Weyl fermions:
\begin{equation}
\bar \psi^i_L \cdot \psi^l_R \psi^i_L\hskip0.5cm {\rm and}\hskip0.5cm 
\bar \psi^l_R \cdot \psi^i_L \psi^l_R,
\label{com}
\end{equation}
respectively to 
build two massive Dirac fermions $\psi^l_n$, neutral with respect to 
the gauge transformations of $SU(2)_L$, and $\psi^i_c$ , carrying a 
$SU(2)_L$ charge. Their propagators in the strong coupling limit are 
given by
\begin{eqnarray}
S^{lk}_n(p)={{i\over a} \sum_\mu \gamma^\mu \sin (p_\mu) +M\over {1\over 
a^2}\sum_\mu \sin^2 (p_\mu) + M^2} \delta_{lk};
\label{sn}
\end{eqnarray}
\begin{eqnarray}
S^{ij}_c(p)={{i\over a} \sum_\mu \gamma^\mu \sin (p_\mu ) + M \over {1\over 
a^2}\sum_\mu \sin^2 (p_\mu ) + M^2}\delta_{ij},
\label{sc}
\end {eqnarray}
where the chiral invariant mass is given,
\begin{equation}
M= 16 {g_1\over a}(1 + {g_2\over g_1} w(p)^2);\hskip0.3cm 
w(p) = \sum_\mu (1-\cos (p_\mu)).
\label{m1}
\end{equation}
These propagator show that in this region not only the ``doublers'' have 
acquired a $O({1\over a})$ mass but also the ``normal'' modes 
($\Pi_a=0$). And this is clearly not acceptable.

Let us now discuss the phase structure of the theory in the neighbourhood
of the critical value $g^c_1$ (while keeping $g_2$ very small). Such 
critical value is determined \cite{gpr} by the propagators of the $8$ composite 
scalars
\begin{eqnarray}
A^{il}_1={1\over \sqrt{2} }(\bar \psi^i_L \cdot \psi^l_R +\bar \psi^l_R 
\cdot \psi^i_L), \hskip1cm 
A^{il}_2={i\over \sqrt{2} }(\bar \psi^i_L \cdot \psi^l_R -\bar \psi^l_R 
\cdot \psi^i_L),
\label{ail}
\end {eqnarray}
corresponding to the real and imaginary parts of the complex fields 
$\bar \psi^l_R \cdot \psi^i_L$. In the strong coupling limit one 
easily obtains 
\begin{equation}
G^{il,jk}_{1,2}(q) = \delta_{ij} 
\delta_{lk} {1\over {4\over a^2}\sum_\mu\sin^2{q_\mu\over 2}
+\mu^2},
\end {equation}
where
\begin {eqnarray}
\mu^2={16 \over a^2} g_1 \left(1+{4g_2 \over g_1 }w(p) w(p^\prime 
)\right)-{8\over a^2 },  
\label{mu}
\end {eqnarray}
and $p,p^\prime$ are external momenta. It is only when the composites
(\ref{ail}) become bound states, i.e. when $\mu^2>0$, that one obtains what we
might call the EP-phenomenon, leading to the appearance of the chiral symmetric
phase (that we call EP-phase) where Dirac fermions propagate according to
(\ref{sn}) and (\ref{sc}). The critical value $g_1^c$ can be determined by 
setting
$\mu^2=0$ Thus, due to the momentum dependence (through the $g_2$-term) of
$\mu^2$, for $g_2\neq 0$ we shall have different critical coupling constants
$g^c_1$ for different ``doublers'', in particular in the strong coupling  limit
for the ``normal'' mode one has $g^c_{1 normal}=0.5$. Thus in the
($g_1,g_2$)-plane from the point ($g^c_{1 normal},0$) there start (See Fig.1)
different critical lines, one for each $\Pi_A$, separating the symmetric
EP-phase from another phase, that Eichten and Preskill expected to be a weak
coupling symmetric phase. Thus, one can define continuum chiral gauge theories
on one of these critical lines where all doublers are decoupled, chiral
fermions remain and gauge symmetry is not broken. We call this expectation as
the EP scenario. However, as pointed out by ref.\cite{gpr}, this EP scenario fails
owing to the fact that critical lines $g^c_1(g_2^c)$ do not separate two
symmetric phases and there is a spontaneous symmetry breaking phase, which we
shall call the NJL-phase, in between the two symmetric phases. This can be quickly
figured out by looking at what one gets $\mu^2>0$ (symmetric phase), 
$\mu^2=0$ (critical lines) and $\mu^2<0$ (broken phase). 

In order to check this latter statement, let us return to the weak 
coupling $g_1 ,g_2$ limit. Based on the analysis of the large-$N_f$ 
($N_f$ is the multiplicity associated with an extra fermionic index, 
e.g. $N_{color}$) weak coupling expansion, we argue that the 
NJL-terms in the action (\ref{action}) induce a spontaneous chiral symmetry 
breaking \cite{njl}. Indeed in the new vacuum the Weyl fields 
$\psi^i_L$ and $\psi^i_R$ pair up to become massive Dirac fields that 
violate $SU(2)$-chiral symmetry, a fact that can be ascertained in the 
following way. The inverse propagator of this massive Dirac fermion 
can be written as 
\begin{eqnarray}
S^{-1}(p)=\left(\matrix{&P_L{i\over a}\sum_\mu \gamma_\mu f^\mu_L 
(p)P_L \delta_{ij} &P_L\Sigma^{il}(p)P_R\cr
&P_R\Sigma^{jk}(p)P_L&P_R{i\over a}\sum_\mu\gamma_\mu f^\mu_R(p)P_R 
\delta_{lk}}\right),
\label{sb}
\end{eqnarray}
where the fermion self-energy function, in the $N_f\rightarrow
\infty $ limit, obeys the ``gap-equation'' 
$(\Sigma^{il}(p)=\delta^{il}\Sigma (p))$:
\begin{eqnarray}
\Sigma(p)=2\int_q{\Sigma(q)\over\den(q)}\left(\tilde g_1+\tilde g_2 w(p) 
w(q)\right),
\label{se}
\end{eqnarray}
where 
\begin{equation}
\int_q\equiv \int_\pi^\pi{d^4q\over (2\pi)^4},\hskip0.3cm \den(q)\equiv 
\sum_\rho\sin^2(q_\rho) +(a\Sigma(q))^2,
\label{int}
\end{equation}
and $\tilde g_{1,2}\equiv g_{1,2}N_f$. Using the parametrization\cite{gpr} 
\begin{equation}
\Sigma (p)=\Sigma 
(0)+\tilde
g_2 v w(p);\hskip0.5cm \Sigma (0) =\rho v,
\label{para}
\end{equation}
where $\rho$ depends on 
$\tilde 
g_{1,2}$ only, we can solve the gap-equation (\ref{se}) in a 
straightforward manner. For $v =O({1\over a})$, one gets 
\begin{eqnarray} 
\rho={\tilde g_1\tilde g_2 I_1\over 1-\tilde g_1 I_\circ}, \hskip1.5cm
\rho={1-4\tilde g_2 I_2\over 4 I_1},
\label{roa}
\end{eqnarray}
where 
\begin{equation}
I_n=4\int_q{w(q)^n\over \den(q)}.
\label{in}
\end{equation}
For $g_1=0$ Eq.(\ref{roa}) implies:
\begin{equation}
\rho=0,\hskip0.5cm {\rm i.e.}\hskip0.5cm \Sigma(0)=0.
\label{hope}
\end{equation}
This means 
that on the line 
$g_1=0$ the ``normal'' modes $(\Pi_A=0)$ remain massless, while all the 
``doublers'' acquire a mass $O({1\over a})$, thus violating chiral 
invariance through the $g_2$ coupling alone.

As for the function $f^\mu _L (p)$ $(f^\mu _R(p))$ in Eq.(\ref{sb}), 
the 
$\psi_R$-shift symmetry for $g_1\rightarrow 0$ allows us to derive for 
$f^\mu _R(p)$ the form 
\begin{equation}
f^\mu_R (p)=\sin (p_\mu),
\end{equation}
while for $f^\mu _L(p)$ in the large $N_f$-limit one obtains:
\begin{equation}
f^\mu_L (p)=Z_2(p)\sin (p_\mu),
\end{equation}
with the wave-function renormalization $Z_2(p=\tilde{p}+\Pi)=const.$\footnote 
{The details of the calculation will be presented elsewhere.} Thus the massive
Dirac 
inverse propagator can be written for both members of the $SU(2)$ doublet
\begin{eqnarray}
S^{-1}(p)={i\over a}\sum_\mu \gamma_\mu \sin (p^\mu) Z_2 
(p)P_L+{i\over a}\sum_\mu  \gamma_\mu \sin (p^\mu) P_R+\Sigma(p),
\label{smu}
\end{eqnarray}
showing that the $SU(2)_L \times SU(2)_R$ symmetry is spontaneously 
broken to $SU(2)$, giving rise to three Goldstone bosons and one massive 
$O({1\over a})$ Higgs mode, as discussed in Ref.\cite{xue}.

For $v\rightarrow 0$ Eq (\ref{roa}) yields a critical line $(\tilde 
g^c_1,\tilde g^c_2)$ separating the NJL-phase from the weak coupling 
symmetric phase, which goes from $(0.4,0)$ to $(0,0.006)$ as indicated in Fig.1. 
Thus the NJL-phase stands between the chiral symmetric doubled weak 
coupling phase and the chiral symmetric EP-phase. 
This fact, that according to our analysis holds even for $N_f =1$, 
shows that the EP-scenario for getting around the no-go theorem, 
unfortunately cannot be realized. Indeed, such scenario 
contemplates the contiguity of the EP and the symmetric weak phase, so 
that there would exist a chiral line (See Fig.1) where the 
theory is chirally symmetric and {\em all } the unwanted doublers get a mass 
$O({1\over a})$. Our analysis shows instead that the EP-phase is 
contiguous to the NJL-phase, that we have just described and agrees
with analysis of ref.\cite{gpr}.

Unfortunately also the NJL-phase is physically not entirely acceptable, 
for even when $g_1 =0$ the $SU(2)_L \times SU(2)_R$ chiral symmetry is 
violated by a 
hard breaking Wilson term \cite{wilson} (See eq. $(14)$), corresponding to 
a dimension $5$-operator. As a consequence, in the well known phenomenon 
that mixes the gauge bosons and the Goldstone modes one expects the gauge 
bosons to acquire a mass $O({1\over a})$, contrary to physics 
expectations.

\vskip1cm
\noindent
{\bf 3.\hskip0.3cm}
In our investigation we are finally left with another strong coupling 
region $(g_1 \ll 1, g_2 \gg 1)$ (See Fig.1) where the $\psi_R$-shift 
symmetry is exact. This region can be 
treated exactly as the one dealt with above. We find that the 
vacuum of the theory retains the $SU(2)_L \times SU(2)_R$ symmetry, and 
its excitation spectrum contains both chiral invariant $(\psi^l_n)$ and 
chiral $(\psi^i_c)$ massive Dirac fermions, whose propagators in 
the strong coupling limit $g_2 \gg 1$ can be determined to be of
the same form as eqs.(\ref{sn},\ref{sc}). The chiral invariant mass, 
instead of eq.(\ref{m1}), this time is given by:
\begin{eqnarray}
M^2= {16\over a^2} g^2_2 \left(w(p)^2+ {g_1\over g_2}\right)^2.
\label{emme}
\end{eqnarray}

As for the charged fermion field $\psi^i_c$, even when 
$g_1\ll 0$ and $g_2\gg 0$, the low energy spectrum still comprises the 
massive Dirac 
mode built by the ``normal'' modes of $\psi^i_L$ and the composite 
$(\bar \psi^k_R \cdot \psi^i_L)\psi^k_R$, a situation that is 
phenomenologically undesirable. One can solve this problem if in the 
$(g_1,g_2)$-plane there existed a wedge where only the unwanted 
doublers are in the EP-phase, while the ``normal'' modes would still 
be massless (does not undergo the NJL spontaneous symmetry breaking) and 
live upon a chirally symmetric vacuum. In order to see 
whether such 
circumstance really occurs, let us proceed as above. We consider again 
the scalars $A^{il}_{1,2}$ (See Eq.(\ref{ail})) and their propagators, whose 
mass $\mu^2$ this time is not given (approximately) by (\ref{mu}), but by:
\begin{eqnarray}
\mu^2= {16\over a^2}g_2 \left( 4w(p) w(p')+{g_1\over g_2} \right) - {8\over 
a^2}.
\label{mub}
\end{eqnarray}
The critical lines for different doublers acquiring chiral invariant masses
are determined, as above, 
by the condition $\mu^2=0$; thus the first ``threshold'' (See Fig.1) is 
encountered when $g_1=0$ at $g_2=0.002$, for the doubler:
\begin{equation}
p=p\prime =(\pi,\pi,\pi,\pi),
\label{d1}
\end{equation}
while the last threshold appears for 
$g_2=0.03$, corresponding to the four doublers with 
\begin{equation}
p=p\prime =(\pi,0,0,0),\hskip0.3cm (0,\pi,0,0),....
\label{d4}
\end{equation}
Thus in  the strong coupling 
approximation for $g_2>0.03$ {\em all} excitations but the ``normal'' 
modes $(p,p\prime =\tilde p)$, i.e. {\em all} doublers, acquire 
chiral-invariant $O({1\over a})$ masses (\ref{sn},\ref{sc},\ref{emme}), 
while the latter modes remain 
massless, on a chirally invariant vacuum, so long as $g_1=0$. 

In order to check latter statement that normal mode remains massless, we
go back to eq.(\ref{hope}) that shows normal mode is massless when the coupling
$g_1=0$, even
for very large value of the coupling $g_2$. This is actually resulted by the 
exact shift-symmetry (\ref{shift}) when the coupling $g_1=0$. Thus, the failure
of the EP-scenario, due to very existence of the NJL broken phase in between
two symmetric phases discussed in our previous section and ref.\cite{gpr} can
be cured when $g_1=0$, where normal modes always remain in the
chirally symmetric phase and we can send $g_2\rightarrow\infty$ to guarantee
that all the doublers are chirally invariantly decoupled. Note that such
an important feature is absent, when $g_1=0$, in the original EP's model\cite{ep}.   

To summarize the above discussion, let us make a journey in the $(g_1,g_2)$-plane 
taking off from the origin, and identify the different phases of the 
theory described by the lattice action (\ref{action}). We walk first in the 
weak 
coupling symmetric phase, populated by massless, doubled fermions: a 
strange, alien world. Diffusing away from the origin we hit a line, 
the NJL-line, that pushes the doublers to the cut-off, i.e. gives them 
a chirally variant $O({1\over a})$ mass, but unfortunately does so also to the chiral 
gauge bosons due to the hard (dimension-$5$) Wilson term, that 
possesses 
now a non-trivial expectation value in the new NJL-vacuum. Thus, 
again, we are in a world completely different from our own. Continuing 
to move out, always keeping $g_1<0.5$, we hit several other 
critical lines where different doublers acquire a chirally invariant $O({1\over a})$ 
mass, up to the last line where the four doublers physical with $|p_\mu |=\pi 
$ get their chiral invariant $O({1\over a})$ masses 
(\ref{sn},\ref{sc},\ref{emme}). Past that line and go down onto $g_1=0$,
we are in the sought 
``physical wedge'' (See Fig.~1): 
here the low energy spectrum is undoubled, it only contains ``normal'' 
modes and, most importantly, the original chiral $SU(2)_L$-gauge 
symmetry for $|p_\mu |\ll 1$ is exact when  $g_1=0$. More precisely, in this 
wedge the ``normal'' mode of $\psi^i_L$, due to the vanishing of $\Sigma(0)$
(\ref{hope}), 
interacts in a completely chiral invariant way. From Eq.(\ref{smu}) 
we can write the propagators of such massless modes as:
\begin{eqnarray}
S_L(p)^{-1}_{ij}= i\gamma_\mu \tilde p^\mu \tilde Z_2 P_L 
\delta_{ij},
\label{sldp}
\end{eqnarray}
and
\begin{eqnarray}
S_R(p)^{-1}_{lk}= i\gamma_\mu \tilde p^\mu P_R \delta_{lk},
\label{splk}
\end{eqnarray}
where $\tilde Z_2$ is the wave-function renormalization for 
$p=\tilde p$ and $g_1=0$.
Thus we see that the scenario envisaged by Eichten and Preskill can be 
finally realized, but only in the physical wedge $g_1 \rightarrow 0$,  $g_2>0.03$. We should also stress at 
this point that, even though our arguments are 
based on various kinds of approximations, strong coupling and large 
$N_f$, there are absolutely no reasons to expect that more accurate 
calculations will change {\em qualitatively} the whole picture. 

This shows that there exists a 
well defined, concrete way to get around the ``no-go theorem'' of 
Nielsen and Ninomiya that has so far barred the road to viable chiral 
gauge theories. The way around the ``no-go theorem'' requires the 
extension of the simple Wilson gauge lagrangian through the addition 
of two quadrilinear fermionic interactions terms (See Eq.(1)), which 
in a particular physical wedge of the $(g_1,g_2)$-plane give 
rise to a ground state where the unwanted doublers acquire a large
chiral invariant $O({1\over a})$ mass by the EP mechanism, while the ``normal'' (low 
momenta) fermionic modes do not get their masses through a Nambu-Jona Lasinio 
spontaneous symmetry breaking mechanism, thus keeping the low-momentum 
interaction chirally invariant. This provides a physical arena to realize
the scenario of the RA.
The theory defined by the action Eq.(\ref{action}) upon the vacuum of the 
``physical wedge'' has all the features of the target theory of the RA. 
As a consequence it can be looked at as a concrete justification of 
the feasibility of the perturbative strategy envisaged in the RA. However,
instead of utilizing in the ``renormalized'' 
perturbation theory the gauge non-invariant Wilson propagator to remove
doublers and the gauge variant counterterms to enforce the chiral gauge symmetries,
one has now in the ``physical wedge'' an undoubled low-energy
effective theory with the gauge invariant propagators (\ref{sn}), (\ref{sc}) 
for doublers and 
(\ref{sldp}), (\ref{splk}) for normal modes respectively, thus allowing 
gauge-invariant counterterms only, as expected in gauge invariant Quantum
Field Theories like QCD and QED.

\vspace*{1cm}
\section*{Figure Captions} 
 
\noindent {\bf Figure 1}: \hspace*{0.2cm} 
The phase diagram for the theory (\ref{action}) in the $g_1-g_2$
plane (at the gauge coupling $g\simeq 0$).

\end{document}